\documentclass{aastex}
\usepackage{emulateapj5}

\newcommand{\n}{\nodata}

\def\bi{\begin{itemize}}
\def\ei{\end{itemize}}

\def\Msun{{\rm\,M_\odot}}
\def\gtrsim{\mathrel{\hbox{\rlap{\hbox{\lower4pt\hbox{$\sim$}}}\hbox{$>$}}}}
\def\lesssim{\mathrel{\hbox{\rlap{\hbox{\lower4pt\hbox{$\sim$}}}\hbox{$<$}}}}

\shortauthors{Lister et al.}
\shorttitle{4C~+12.50: A Superluminal Precessing Jet}

\begin{document}
\title{4C~+12.50: A Superluminal Precessing Jet in the Recent \\
Merger System IRAS 13451+1232}

\author{M. L. Lister\altaffilmark{1}, K. I. Kellermann\altaffilmark{1},
R. C. Vermeulen\altaffilmark{2}, M. H. Cohen\altaffilmark{3},
J. A. Zensus\altaffilmark{4}, E. Ros\altaffilmark{4}}

\altaffiltext{1}{National Radio Astronomy Observatory, 520 Edgemont
Road, Charlottesville, VA 22903--2454}

\altaffiltext{2}{Netherlands Foundation for Research in Astronomy, Postbus 2, 7990 AA 
Dwingeloo, The Netherlands}
\altaffiltext{3}{California Institute of Technology, Department of Astronomy, MS 105--24, 
Pasadena, CA 91125, USA}
\altaffiltext{4}{Max-Planck-Institut f\"ur Radioastronomie, Auf dem H\"ugel 69, D--53121 
Bonn, Germany}

\email{mlister@nrao.edu}

\begin{abstract}
We present the results of a multi-epoch VLBA study of the powerful
radio source 4C~+12.50 (PKS 1345+125) at a wavelength of 2 cm. This
compact radio source is associated with a hidden quasar whose host
galaxy shows signs of a recent merger. It has been classified as a
compact symmetric source (CSO) due to its small overall size ($\sim
220$ pc) and twin-jet morphology, although it also has faint extended
emission that may be a relic of previous activity. We report the
detection of exceedingly high linear fractional polarization in
isolated features of the southern jet (up to 60 \%), which is highly
unusual for a CSO. Given the large amount of gas present in the host
galaxy, we would expect significant Faraday depolarization across the
whole source, unless the depolarizing gas is fairly clumpy.  The
southern jet also contains two features that are moving outward from
the core at
apparent speeds of $v/c = 1.0 \pm 0.3$ and $1.2 \pm 0.3$. These
represent the first positive detections of superluminal motion in a
CSO, and taken together with the jet to counter-jet flux ratio, suggest
an intrinsic flow speed of $v/c = 0.84 \pm 0.12$. The apparent ridgeline
of the jet and counter-jet are consistent with a conical helix of
wavelength 280 pc that is the result of Kelvin-Helmholtz instabilities
driven by a slow precession of the jet nozzle. A fit to the data
implies that the nozzle is precessing around a cone with half-angle
$23 \arcdeg$, whose axis lies at an angle of $82 \arcdeg$ to the line
of sight. We suggest that the ``S''-shaped jet morphologies commonly
seen in recent AGN outflows such as 4C~+12.50 may simply reflect the
fact that their black hole spin axes are still precessing, and have
not had sufficient time to align with their accretion disks.

\end{abstract}


\section{Introduction}

A significant fraction of all active galactic nuclei found in high
frequency radio surveys is made up of gigahertz
peaked-spectrum (GPS) and compact steep-spectrum (CSS) sources. These
AGN have convex radio spectra that peak at rest-frame frequencies of
up to 50 GHz \citep{EJS96}. There is a well known correlation between
peak frequency ($\nu_m$) and overall source size in the radio ($\nu_m
\propto l^{-0.65}$; \citealt*{O98}) that is likely an evolutionary effect
associated with either free-free \citep{BDO97} or synchrotron
self-absorption \citep{OB97}. The historical division between GPS and
CSS classifications has been somewhat arbitrary, with the latter
having observed peak frequencies that usually lie below $\sim 500$
MHz, and are therefore difficult to measure. As a consequence of the
$\nu_m$--size correlation, CSS typically have sizes larger than 1 kpc,
which is sometimes used as a canonical value for dividing CSS and GPS
objects.

There has been much recent work involving a particular sub-class of
GPS sources that show two-sided radio jets on parsec scales.  Their
radio morphology is dominated by two ``mini-lobes'' which are
typically separating at $\sim 0.3 \;c$ 
\citep{PCO02}. The core components are usually weak or not
detected in VLBI images (with some exceptions; see, e.g.,
\citealt*{AWB98}). Their subluminal expansion speeds, relatively
stable flux densities, and two-sided morphologies suggest that their
jets lie close to the plane of the sky, and that their high apparent
luminosities ($P_{1.4\;
\mathrm{GHz}} \simeq 10^{25} \; \mathrm{W\, Hz^{-1}}$) are not the
result of relativistic beaming.  Although these AGN are usually
referred to as ``compact symmetric objects'' (CSOs), their
parsec-scale jets are rarely symmetric, and in some cases are
associated with much larger radio sources that extend out to many
kiloparsecs \citep{SBO90,SOD98}. In such sources (e.g., 0108+388;
\citealt*{BOB90}), it has been speculated that the CSO represents a
renewed period of activity in the life cycle of the AGN, since the
kinematic age of the CSO ($\lesssim 1000$ years; \citealt*{OCP99}) is much
smaller than that inferred for the large-scale structure.

If CSOs are indeed radio sources that have only recently become
(re)active, it should still be possible to find traces of the
triggering events in their host galaxies. The current paradigm for
relativistic jet formation in AGN is that a merger disturbs the
gas in the inner regions of the host galaxy, sending it toward the
central black hole. An accretion disk forms and then provides the fuel
for a relativistic outflow. Although this scheme is now widely
accepted, there are relatively few nearby AGN in which the
predictions can be examined in detail. One such system is the
ultra-luminous infrared galaxy (ULIG) IRAS 13451+1232, which contains
a double nucleus and appears to have undergone a recent merger. What
makes this AGN particularly useful for investigating both the merger
and CSO ``youth'' scenarios is that it harbors the powerful radio
source 4C~+12.50, which is one of the closest known GPS/CSS
sources. Although 4C~+12.50 has a CSO-type morphology on
parsec-scales, it also has weak, extended radio emission which is
possibly a relic of previous nuclear activity.

Here we present multi-epoch VLBA\footnote{The Very Long Baseline
Array (VLBA) is a facility of the National Radio Astronomy
Observatory, operated by Associated Universities Inc., under
cooperative agreement with the National Science Foundation.} 
observations of 4C~+12.50 which reveal for the first time both the
existence of superluminal motion and high linear polarization in the
jets of a CSO.  We find evidence that suggests that the jet nozzle
of 4C~+12.50 is slowly precessing with time. We discuss whether the
twisted, symmetric jet morphologies seen in some CSOs may simply
reflect the fact that their black hole spin axes are still precessing,
and have not had sufficient time to reach a stable orientation.

In \S~2 we describe
our observations and data reduction. We discuss the overall
host galaxy and radio properties of 4C~+12.50 in \S\S3.1 and
3.2, respectively. In \S~4 we examine the
proper motions in the jet, and discuss the constraints on the viewing
angle and flow speed in \S~4.1. In \S~5
we develop a helical precession model that accounts for the dramatic
jet curvature in the source, and consider the evidence for 4C~+12.50
being a young radio source in \S~6. We summarize our
findings in \S~7.

In this paper we assume a standard cosmology with $\Omega_m =\ 0.3$,
$\Omega_\Lambda = 0.7$, and $H_o = 70 \; \rm km\; s^{-1} \;
Mpc^{-1}$. The redshift of IRAS 13451+1232 ($z = 0.122$) implies a
luminosity distance of 570 Mpc, and an angular size of 1 mas on the
sky corresponds to 2.20 pc.


\section{\label{observations} Observations and data reduction}

In order to study the jet kinematics of 4C~+12.50, we have
used five individual epochs of 2 cm VLBA data from three separate
observing programs that spanned the time range 1996 to 2001. The first
epoch consists of archival VLBA data observed by
\cite{SDO01}. The next three epochs are from the VLBA 2 cm
survey \citep{KVZ98,ZRK02}, and the final epoch is from a full-track
observation carried out specifically for this study. The first and
last epochs were recorded in dual circular polarization, and included
a single VLA antenna in addition to the ten antennas of the VLBA (see
Table~\ref{journal}).

The data from the individual epochs were all correlated with the VLBA
correlator in Socorro, NM and were processed in identical fashion
using AIPS and Difmap.  We applied a-priori
amplitude corrections using antenna gains and system temperatures
measured during the run, but did not perform any atmospheric opacity
corrections, as the antenna gains were adequately calibrated in
subsequent self-calibration and imaging iterations in Difmap. 

We determined the antenna polarization leakage factors for the final
epoch by running the AIPS task LPCAL on the calibrator sources OQ~208
(1404+286) and PKS 1413+135. These sources were observed at regular
intervals during the run in order to obtain a large range of
parallactic angles. The leakage factors ranged up to $6\%$, with
typical values of $\sim 2\%$. By comparing the results obtained from
both calibrators, we estimate that our leakage factor solutions are
accurate to within $\sim 0.1\%$. Our re-analysis of the first epoch
data observed by \cite{SDO01} using the sources OQ~208,
0738+313, 0742+103, and 0743--006 gave similar results. 

We established the absolute electric vector position angle (EVPA) on
the sky for the last epoch using five scans of the calibrator
3C~279 separated in hour angle throughout the observing run. The
integrated EVPA of 3C~279 was measured at 1.3, 3.6, and 6 cm on 2000 Dec
28 with the VLA as part of the VLA/VLBA Polarization Monitoring Program
\citep{TM00}. By fitting a rotation measure of 54.7 $\rm rad/m^2$ to
these data, we interpolated an EVPA of 51.6$\arcdeg$ at 2 cm for this
epoch. We then rotated the VLBA EVPAs so that in they were in
agreement with this value. Based on previous applications of this
technique (e.g., \citealt{LS00, L01}), we estimate that our EVPAs are
accurate to within $\sim 5 \arcdeg$.  


We calibrated the EVPAs of the first epoch by comparing single-dish
measurements of the compact sources 0552+398 and 2134+004 made at the
University of Michigan Observatory (UMRAO) to our integrated VLBA
EVPAs. A rotation of $-17$ degrees was needed to bring the VLBA EVPAs
into agreement with the Michigan data. A comparison of our 1996 and
2001 images shows no changes in polarization structure between these
epochs, with EVPAs differing by less than $\sim$2 degrees.

\section{Discussion}

In this section we discuss previous observations of 4C~+12.50, as well
as those of its host galaxy IRAS 13451+1232. We describe the results
of our VLBA observations in \S~\ref{masmorph}.

\subsection{Host galaxy properties\label{host}}

IRAS 13451+1232 has been imaged extensively by space- and ground-based
observatories in the optical and infrared (e.g., \citealt{Sco00}).
Early optical images by \cite{GS86} revealed two nuclei separated by
1.6 arcsec (3.5 kpc), and a V-band image by \cite{H86} showed
irregular isophotes, several nearby small companion galaxies, and a
strongly curved tidal tail. The disturbed features of this system and
the relatively large nuclear separation are indicative of a recent
merger event involving at least one spiral, gas-rich galaxy
\citep{GS86}. The east nucleus is optically dull and appears as
reddened starlight. The northwest nucleus, which is coincident with
the radio source \citep{ACF00}, has a Seyfert 2 optical spectrum and
is highly polarized in the ultraviolet ($m =16 \%$;
\citealt*{HAC99}). It is highly reddened ($B - I = 2.5$; \citealt*{SS00}) with an
estimated $A_V$ of 3.5 \citep{K95}, and has broad Pa($\alpha$)
emission in the infrared \citep{VSK97}. These observations suggest the
presence of a quasar buried behind a foreground dust screen, which is
further supported by the detection of hard X-rays from this source by
ASCA (\citealt{IU99,ODW00}).

The likely source of fuel for the quasar is a vast quantity of
molecular gas that appears to be left over from the merger. The
northwest nucleus marks the peak of strong compact CO ($1 \rightarrow
0$) emission at 2.7 mm, which \cite{EKM99} found to be concentrated to
within the inner 2 kpc of the galaxy.  In addition to the CO
emission, \cite{MSK89} detected HI absorption in front of 4C~+12.50.  The HI and CO profiles span the same velocity range and are
likely spatially coincident. \cite{MSK89} derive a molecular gas mass
of $6.5 \times 10^{10} \Msun$, which is among the highest ever
measured for an ULIG.

\subsection{Radio properties of 4C~+12.50 \label{radioprops}}
4C~+12.50 is a relatively powerful radio source, with  $L_{\rm 408\; MHz} = 2.7
\times 10^{26} \rm \; W\; Hz^{-1}$. It has a curved radio spectrum 
with a spectral index of approximately $\alpha = -0.45 \; (S_\nu
\propto \nu^\alpha)$ at frequencies
between 0.4 and 30 GHz.  There is a sharp cutoff at low frequency
($\sim 400$ MHz) characteristic of free-free or synchrotron self
absorption.  Its radio flux density is relatively stable at 318 and
430 MHz; \cite{SAG99} report only very weak ($\sim 1\%$) variations over a
14 year time interval. Long term 8 GHz monitoring at UMRAO\footnote{http://www.astro.lsa.umich.edu/obs/radiotel/umrao.html} showed a
slow rise of 300 mJy over several years in the early 1980s (also seen at 5 GHz by Bennett
et al. 1984), followed by a period of relatively constant flux density
until the end of monitoring in 1992. During the rise period the $5-8$
GHz spectral index was considerably flatter than at other times, which
was likely due to a brightening of the flat spectrum core (see
\S~\ref{masmorph}).  When UMRAO observations resumed in 2000, the 8
GHz flux density had returned to its original 1980 level.

\subsubsection{Arcsecond-scale morphology}
The radio morphology of 4C~+12.50 is generally compact, with 18 and
6 cm VLA A- and B-array observations showing that most of the flux
density comes from a region $\sim 0.1$ arcsec (220 pc) in extent
\citep{SMC89, A85, UJP81}. This size has been confirmed with MERLIN at
18 cm \citep{XSD02}. The total interferometric flux densities at 18 cm
are generally consistent with single-dish measurements (e.g.,
\citealt{MDC94}), suggesting that the compact component accounts for
nearly all of the cm-wave flux density. However, a VLA D-array image
at 6 cm by \cite{CB91} does show two very weak (a few mJy) extensions
of diffuse emission, one north of the core, and one to the
southwest. An extracted 20 cm image from the VLA FIRST survey \footnote{http://sundog.stsci.edu} also shows slightly extensions in these directions. The total extent of the
diffuse emission is approximately 20 arcsec (44 kpc).  Without more
sensitive, high-resolution observations it is difficult to speculate
whether or not this represents material emitted during an earlier
period of AGN activity in the system. We note that the spatial extent
of 4C~+12.50 is similar to that of 0108+388 (30 kpc; \citealt*{BOB90}).

\subsubsection{Milliarcsecond-scale morphology\label{masmorph}} 
\begin{figure*}
\epsscale{0.7}
\plotone{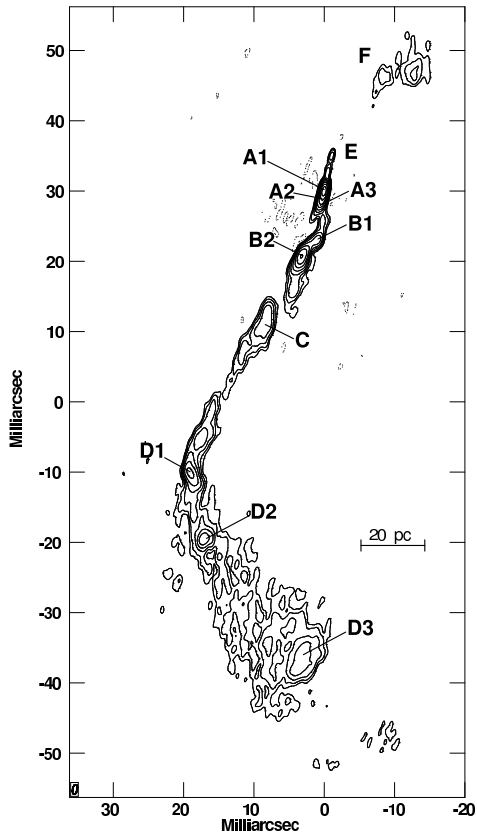}
\caption{\label{fullmap} Naturally-weighted total intensity 2 cm VLBA image
of 4C~+12.50 at epoch 2001.01. The restoring beam dimensions are
$1.41 \times 0.63$ milliarcsec, with major axis at position angle $-10 \arcdeg$. The
contour levels are $-0.4$, 0.4, 0.8, 1.6, 3.2, 6.4, 12.8, 25.6, 51.2,
and 102.4 $\rm mJy \; beam^{-1}$. }
\end{figure*}

The compact structure in 4C~+12.50 has been imaged with VLBI at
wavelengths ranging from 2 to 18 cm \citep{FCF96, SOB97,FFP00, SDO01,
ZRK02, XSD02}. In the following discussion we will use, as much as
possible, the same component nomenclature as \cite{SOB97}. In
Figure~\ref{fullmap} we show our full-track 2 cm VLBA image (epoch
2001.01), which has a significantly lower noise level ($90 \; \mu
\rm Jy \; beam^{-1}$) than the epoch 1996.30 image of \cite{SDO01},
due to the larger bandwidth and longer integration time (see
Table~\ref{journal}).  The overall structure consists of a compact
core component ({\bf A1}) associated with the active nucleus, a
well-defined jet extending 75 mas to the south, a faint extension to
the north of the core, and a more diffuse region to the northwest
({\bf F}).

\cite{SDO01} and \cite{XSD02} both identified component {\bf A1} with the
core on the basis of its relatively flat spectral index between 2 and
18 cm. We find additional evidence supporting this identification, in
that it is the brightest, most compact feature in our image. In
blazars, the core is usually appreciably polarized (e.g.,
\citealt*{L01}), and indeed {\bf A1} is the only polarized feature in
this region of the source (see Fig.~\ref{corezoom}). The relatively
high brightness of the core compared to the lobes is typical for
CSOs found in nearby galaxies with large apparent magnitudes
\citep{AWB98}.

The northern jet is significantly fainter and shorter than the
southern jet. It is well-collimated near the core, but then
disappears, only to reappear as a region of diffuse emission further
to the northwest. This emission was also detected at 6 and 18 cm by
\cite{SOB97}  and  \cite{XSD02}, respectively,  leading to the initial
classification of 4C~+12.50 as a CSO.

The southern jet is somewhat knotty, and has an unusually high
length-to-width ratio compared to the jets of typical AGN. For the
purposes of measuring apparent motions, we have modeled it using 8
distinct emission regions labeled {\bf A1} through {\bf F} in
Figure~\ref{fullmap}.  The jet remains narrow and well-collimated
until it undergoes a sharp bend near {\bf D1}. The small opening angle
and knottiness prior to {\bf D1} suggests a high internal Mach number
that is typical of powerful FR-II type jets.  Past {\bf D1} the jet
widens considerably and becomes rather diffuse, with much of the flux being
resolved out in our high resolution image. A simultaneous 2 cm
observation at UMRAO gave a total flux density of 1420 mJy, compared
to only 950 mJy recovered in our VLBA observation (M. and H. Aller, private
communication).  The 6 cm VLBI image of \cite{SOB97} shows that the
southern jet past {\bf D1} is embedded in a roughly circular,
steep-spectrum emission region with diameter $\sim$ 30 mas. Their
image lacks a similar amount of the total flux at 6 cm ($\sim$ 400
mJy).

It is useful to examine whether the increase in apparent opening angle
past {\bf D1} could be a projection effect resulting from bending toward the
line of sight. As we will show in \S~4, the jet flow prior
to the bend is relativistic, and is likely oriented close to the plane
of the sky. The apparent jet opening angle before to the bend is $\Psi_i =
1.1 \arcdeg$, while after the bend it increases to $\Psi_f =
8\arcdeg$. Assuming an initial viewing angle of $i = 90 \arcdeg$, then
the change in viewing angle past the bend is ${\Delta\theta} =
\arcsin{(\tan{\Psi_i} / \tan{\Psi_f})}  $.  This would imply a change of
viewing angle of $\sim 82 \arcdeg$ toward the line of sight. Since
the jet flow is relativistic, beaming effects would be expected to
cause a significant apparent brightening in the jet past {\bf D1}, yet the
surface brightness remains relatively constant before and after the
bend. We therefore conclude that the increase in opening angle past
{\bf D1} is real and not a projection effect. 


\subsubsection{Polarization properties}

The integrated linear polarization of 4C~+12.50 is extremely low.
VLA-A configuration observations of \cite{SOD98} give an upper limit
($m < 0.2 \%$) at wavelengths from 3.6 to 18 cm; similar upper limits
were obtained at 2, 3.6 and 6 cm during the UMRAO monitoring
program. Although \cite{WdeP83} claimed to measure very weak circular
polarization at 49 cm (V = $-0.098$ Jy), there have been no reported
detections of circular polarization at other wavelengths. The
generally low polarizations of CSO and GPS sources have been
attributed to Faraday depolarization, possibly by a foreground screen
associated with the narrow-line region \citep{O98}.

\begin{figure*}
\epsscale{0.85}
\plotone{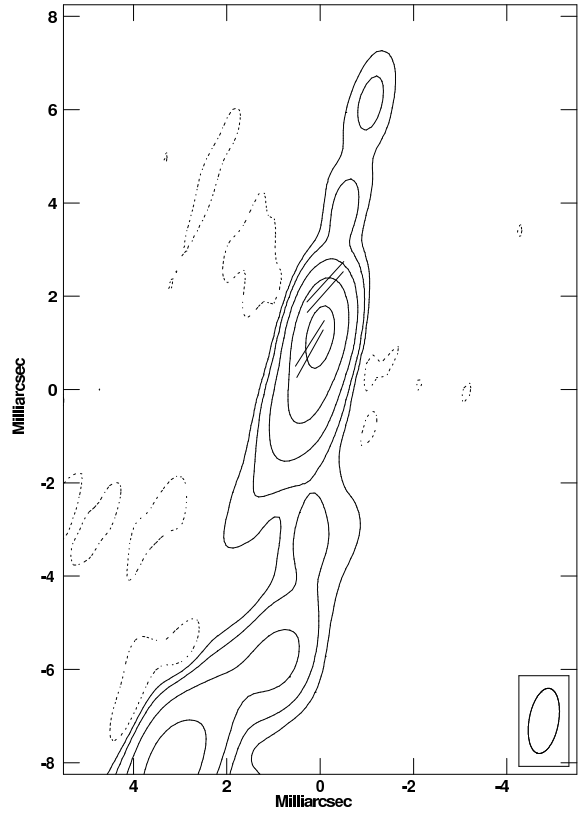}
\caption{\label{corezoom} Close-up view of the core region. The
contours represent total intensity, with levels at $-0.4$, 0.4, 1.6,
6.4, 25.6, and 102.4 $\rm mJy \; beam^{-1}$. The electric vectors are
proportional to linearly polarized flux, with 1 milliarcsec = 303 $\rm
mJy\; beam^{-1}$. }
\end{figure*}

\begin{figure*}
\epsscale{0.95}
\plotone{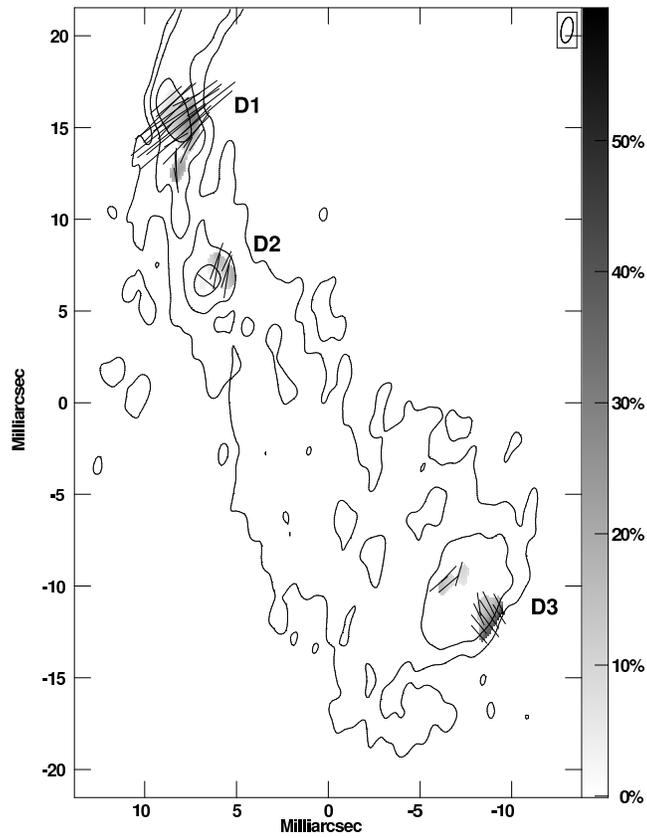}
\caption{\label{southjet} Close-up view of southernmost jet region.  The
contours represent total intensity, with levels at $-0.4$, 0.4, 1.6 and
6.4 $\rm mJy \; beam^{-1}$. The electric vectors are proportional to
linearly polarized flux, with 1 milliarcsec = 303 $\rm mJy\;
beam^{-1}$. The greyscale represents fractional linear polarization in
per cent. }
\end{figure*}

The milliarcsecond-scale linear polarization structure of 4C~+12.50
at epoch 2001.01 is shown in Figures~\ref{corezoom} and
\ref{southjet}.  As expected, the integrated fractional polarization level over the
entire source is very low ($m = 0.7\%$), with the
fractional polarization at the fitted core position being only (0.3
\%). The mean level for the surrounding region is $\sim
0.5\%$. However, surprisingly high fractional polarization levels are
found in the southern region of the jet. Both the total and polarized
intensity contours at {\bf D1} are extended toward the southwest, with
the electric vectors being aligned roughly perpendicular to this
direction.  The fractional polarization gradually increases toward the
southwest, reaching a maximum value of $\sim 30\%$ at the outer
edge. The fractional polarization averaged over the entire feature is
11\%. The polarized emission at {\bf D2} is also concentrated toward
the west, and has mean and peak values of 9 and 20\%,
respectively. The most interesting polarization feature lies at the
extreme southern end of the jet, near component {\bf D3}. Along an
extension of the line connecting components {\bf D2} and {\bf D3}, the
fractional polarization steadily increases to a maximum of $\sim 60\%$
at the outer edge, where both the I and P contours become extremely
steep. The high degree of field order and perpendicular E
vectors of this feature are consistent with a termination shock or
``working surface'' where the jet encounters the interstellar medium
and rapidly decelerates. 

The polarized features at {\bf D1} and {\bf D2} are likely
associated with shocks that energize relativistic particles in the
flow and order the magnetic field. As a jet loses kinetic energy
through a series of internal shocks, its Mach number will drop, which
can eventually lead to subsonic flow and subsequent expansion and
decollimation \citep{ML02}. Such a situation may be present in the
inner kpc-scale jets of FR-I sources such as 3C~31, in which standing
shocks are thought to trigger a sudden flaring in the jet opening
angle \citep{LB02}. The jet of M87 also has a relatively constant
opening angle until it is disrupted at the site of an oblique shock
\citep{BB96}. Such shocks can arise from the well-known helical ($m =
1$) Kelvin-Helmholtz instability produced as the jet propagates
through the interstellar medium.  We will discuss further evidence for
K-H instabilities in the jet of 4C~+12.50 in \S~\ref{helicalstreaming}.

The isolated regions of high polarization seen in 4C+12.50 are highly
unusual for a GPS/CSS source, and are somewhat unexpected, given the
large quantity of gas present in its host galaxy. If sufficiently
magnetized, such a medium would be expected to efficiently depolarize
any synchrotron radiation from the jet. It is possible, however, that
the intervening medium has a low filling factor and is rather
clumpy. Recent Faraday rotation studies (e.g., \citealt*{ZT02}) have
revealed highly non-uniform environments in the inner regions of
AGN. Indeed, there are several isolated regions of the jet (e.g.,
{\bf B1}, {\bf B2}, and {\bf C}) that have comparable brightness to
{\bf D2} and {\bf D3}, but have low fractional polarization, perhaps
as a result of external Faraday depolarization. Also, the sharp
intensity and polarization gradients at {\bf D3} suggest that the jet
has been disrupted after slamming into a high density region.

\section{Jet kinematics\label{kinematics}}

We have investigated proper motions within the radio jet by using the
``modelfit'' task in Difmap to fit elliptical Gaussian components to
isolated bright features (with the exception of component {\bf B1}
where we used a $\delta-$function component). We began by fitting a
model to a single epoch (1999 Nov 1). We then fit the other epochs by
allowing the component fluxes and positions to vary, and keeping all
other parameters fixed. We also experimented using combinations of
circular and $\delta$-function components, and allowing all parameters
to vary, but these methods yielded a much higher degree of scatter in
the component positions across epochs. 

The formal $\chi^2$ of our fits were significantly higher than unity
due to the large amount of diffuse structure present in the jet, which
cannot be easily modeled with Gaussian components.  To obtain error
estimates on the component positions, we used the Difwrap package
\citep{L00}. This program varies the component positions by regular
increments until the model no longer provides an acceptable fit to the
visibilities. We list the fitted component positions and associated
errors for each epoch in Table~\ref{cptpos}. For the bright, isolated
components, our positional errors are typically $\sim 1/6$ the width
of the restoring beam. 

\begin{figure*}
\epsscale{0.75}
\plotone{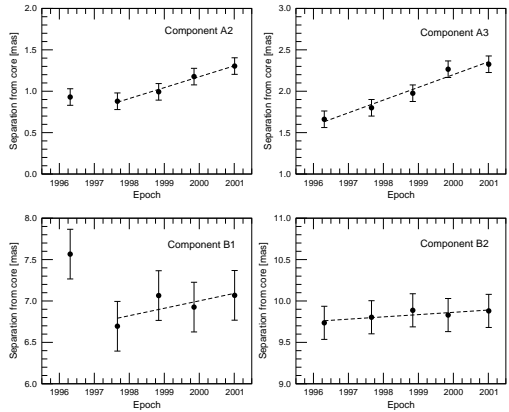}
\caption{\label{ABCD} Plots of separation from core ({\bf A1}) versus time
for components {\bf A2}, {\bf A3}, {\bf B1}, and {\bf B2}. The dotted
lines indicate best-fit linear regressions to the data points. The
fits for components {\bf A2} and {\bf B1} exclude the 1996 epoch.}
\end{figure*}

In Figure~{\ref{ABCD}} we show separation versus time plots for the
four innermost jet components ({\bf A2}, {\bf A3}, {\bf B1}, and {\bf
B2}) with respect to the core component ({\bf A1}), which we consider
to be stationary. With the exception of the first epoch for components
{\bf A2} and {\bf B1}, the component positions are well-described by
linear motion outward from the core. In the case of {\bf A2}, the
fitted position at epoch 1996.30 is likely affected by blending with
the core. The discrepancy in the 1996.30 position of {\bf B1} is much
larger, however, and could be due to a component misidentification at
this epoch. 

We performed linear regression fits to all of the component positions,
the results of which are listed in Table~\ref{cptprops}. The only
components with statistically significant proper motions are {\bf A2}
and {\bf A3}. The remainder are all consistent with zero speed to
within the errors. Our time baseline of 4.7 years allows us to set an 
upper limit of $2\, c$ on their speeds.

The mildly superluminal speeds of {\bf A2} and {\bf A3} ($v/c = 1.0
\pm 0.3$ and $1.2 \pm 0.2$, respectively) indicate that the inner jet
of 4C~+12.50 has only a moderate Lorentz factor and/or is viewed at a
large angle to the line of sight.  Although slow apparent speeds are
possible in a highly relativistic jet that is viewed inside the
critical angle $1/\Gamma$, where $\Gamma$ is the bulk Lorentz factor,
this is not likely to be the case for 4C~+12.50. Its lack of blazar
properties (i.e., strong variability and polarization) and its
two-sided morphology suggest that the jets lie fairly close to the
plane of the sky. Provided that the pattern speed is the same as that
of the bulk flow, the apparent speed of component {\bf A3} limits the
viewing angle in that region of the jet to be within 80 degrees of the
line of sight, and the bulk flow speed to be greater than $0.77\, c$.

Superluminal motion has so far been detected in only a few CSS
sources: 3C 138 \citep{CDF97}, 3C~147 \citep{APK90}, 3C~216
\citep{PFF00}, CTA~102 \citep{JMM01}, and 3C~380  \citep{PW98}. All of
these sources have one-sided jets on parsec scales. To date there have
been no reported superluminal speeds in the jets of CSOs. \cite{TV97}
reported a speed of $1.3 \, c$ for one component (N5) of 1946+708, but
with a high degree of uncertainty due to positional errors associated
with component blending.

\subsection{Constraints on jet speed and viewing angle\label{viewingangles}}

The relative flux densities of the jet and counter-jets in 4C~+12.50
can be used to constrain their speed and orientations, provided that they
have intrinsically identical properties. The overall morphology and
jet ridgeline of the northern counter-jet region ({\bf CJ}) differs
considerably from the southern jet, even taking into account the
possibility of relativistic time-delays. In particular, there is no
diffuse lobe of emission surrounding the northern jet, which suggests
that the external environments at the ends of the northern and southern
jets are considerably different. Any local differences in the external
medium are likely to be less extreme in the small region close to the
central engine, so we have chosen to confine our measurements of the
jet-to-counter-jet flux density ratio ($J$) to within 6 mas of the core.

The integrated flux density of the region $ 1.7 < r < 6$ mas north of
the core is 4 mJy, while the same region south of the core has $S =
31$ mJy, giving a ratio of $J \sim 8$ (we have excluded the region $r
< 1.7$ mas in order to eliminate any contribution of the core flux due
to smearing of the restoring beam along the jet axis).  At the
position of component {\bf A3}, the flux ratio $J$ is approximately
7. We therefore adopt a value of $J = 8 \pm 1$ for the inner jet in the
following analysis.

Assuming Doppler beaming (e.g., \citealt*{UP95}),
$\beta\cos{i} = (J^{1/p} -1)/(J^{1/p} + 1) = 0.37$, where $\beta$ is
the intrinsic jet speed in units of $c$, $i$ is the angle to the line of
sight, and $p = 2 -\alpha$ for continuous jet emission. Here we 
assume a typical value of $\alpha = -0.7$ for the jet spectral index.

\begin{figure*}
\epsscale{0.6}
\plotone{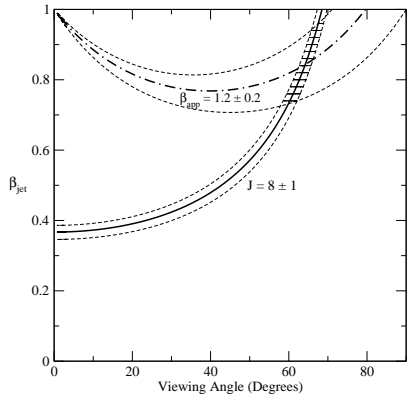}
\caption{\label{viewang} Diagram showing constraints on the intrinsic jet
speed and viewing angle of the southern jet. The hashed region shows
the allowed range of $\beta$ and $i$ given the measured speed 
($\beta_{app} = 1.2$; dot-dashed line) and jet-counter-jet flux ratio ($J
= 8$; solid line) at the position of component {\bf A3}.}
\end{figure*}

If we make a further assumption that the pattern speed of component
{\bf A3} ($\beta_\mathrm{app} = 1.2$ c) is representative of the bulk speed
responsible for the Doppler boosting, then the bulk Lorentz factor is
$\Gamma = (1/2)(J^{1/p}+1)(J^{1/p}-\beta_{app}^2)^{-1/2} = 1.9$, and
$\beta = 0.84 \pm 0.12$. The inferred viewing angle at component {\bf A3} is
$i = 64 \pm 6 \arcdeg$. In Figure~\ref{viewang} we show a graphical
representation of these constraints in the $\beta-i$ plane. The hashed
region shows the allowed range of speed and viewing angle for the jet
at the position of component {\bf A3}. 

The high intrinsic speed of the 4C~+12.50 jet suggests that there are
likely to be other CSOs with end-on orientations that are appreciably
beamed.  Such a population has not yet been positively
identified. However, their high intrinsic luminosities suggest that
they are likely bright enough to already be present in current radio
source catalogs.  Given that the maximum possible Doppler factor of
4C~+12.50 ($\delta
\simeq 2\Gamma \simeq 4$) is much less than those inferred for the
most highly beamed blazars (e.g., $\delta \simeq 80$;
\citealt{FKW99}), the beamed versions of CSOs would be expected to
have less extreme properties. Their integrated fractional
polarizations should be much lower than blazars, due to Faraday
depolarization by dense surrounding gas.

\section{Evidence for jet precession\label{precession}}

The jet ridgeline in 4C~+12.50 traces out an ``$\varepsilon$'' shape
that is highly reminiscent of those seen in galactic microquasars such
as SS 433 \citep{HJ81,SMR81}. The latter have been successfully fit by
jet models in which the nozzle axis precesses slowly and material
flows out on ballistic trajectories.  In these models the jet
ridgeline appears to trace out an apparent three dimensional helical
path on a conical surface, but the velocity vectors of the jet
material are all directed radially outward from the nozzle. In cases
where the velocities are relativistic, the receding jet ridgeline
appears compressed, since we are seeing radiation emitted at an
earlier time than that of the main (forward) jet.

\subsection{Ballistic or streaming motion?}

On first inspection, the morphology of 4C~+12.50 appears to fit the
main predictions of the ballistic model: the counter-jet appears
compressed and has a similar bend to the main jet, and all of the jet
emission is confined within two conical regions whose apexes are
located at the core. Most importantly, the sinusoidal path of the
southern jet ridgeline is consistent with that of a helix seen in
projection. A closer examination of the southern jet, however, reveals
several features which suggest that the jet material is not moving on
simple ballistic trajectories. The most convincing evidence can be found at the
very tip of the southern jet (component {\bf D3} in
Fig.~\ref{southjet}). Both the {\bf E} vectors and the gradients in
total intensity and in fractional polarization run nearly perfectly
parallel to the jet ridgeline that joins {\bf D2} and {\bf D3}. If the
material at {\bf D3} were moving radially outward from the core, one
would have to postulate that all of these alignments occurred simply
by chance, which  is highly unlikely. Also, the fact that the
sites of high fractional polarization are located at the major bend
and terminus of the southern jet cannot be easily
explained by the ballistic model. A more plausible explanation is
that the jet material is streaming parallel to the jet ridgeline, and
undergoes shocks at {\bf D1} and {\bf D3}. Strong evidence for helical
streaming motion has already been found in other AGN jets such as
3C~273 \citep{LZ01} and BL Lacertae \citep{DMM00}.

\subsection{Helical streaming model\label{helicalstreaming}}

Theoretical studies of Kelvin-Helmholtz (K-H) instabilities in
relativistic jets (e.g., \citealt{H87}) have shown that they are
susceptible to large helical twisting patterns provided there is a
suitable perturbation mechanism present at the jet nozzle. These
findings have been confirmed by extensive numerical simulations (e.g.,
\citealt{HCC94}) that show a variety of K-H surface and displacement
modes.  According to the K-H model of \cite{H87}, the n = 1 helical
fundamental mode displaces the entire jet. Higher modes merely produce
helical fluting patterns on the surface, but do not displace the jet
ridgeline.  For the $n = 1$ mode in a conical jet, the dependence of
helix wavelength on distance along the jet axis ($r$) is of the form
$\lambda(r) =
\lambda_1 r^\epsilon$, where $\epsilon$ depends on the external pressure
gradient and the adiabatic index of the jet. The conical boundaries of
the helix imply that the pitch angle remains constant (i.e., if $\epsilon =
1$), or decreases with r (if $\epsilon > 1$). The pitch angle of the jet in
4C~+12.50 appears to {\it increase} down the jet, however, which
would imply $\epsilon = 0$. Such a situation could arise if the instability
is driven by precession and the wave speed remains constant down the
jet (P. Hardee, private communication).

\subsection{Helical model fitting}

A major difficulty in fitting helical models to relativistic jets is that a
minimum of six parameters are needed to describe a helix in
three-dimensional space (see Table~\ref{helixparms}). In  our case,
however, we can narrow the parameter space considerably by using the
constraints on the speed and orientation of the inner jet from
\S~\ref{viewingangles}.  In fitting a conical
helix to the jets of 4C~+12.50, we have chosen to adopt the 
right-hand coordinate system of \cite{GGU82}, in which the observer's
line of sight lies along the {\bf x} axis, and the {\bf y--z}
plane represents the plane of the sky. The helix axis ($\bf z'$) lies in
the {\bf x--z} plane at an angle $i$ to the {\bf x} axis. The helix
has a constant wavelength, and lies on
the surface of a cone of half angle $\psi$ whose axis is coincident
with $\bf z'$. The parametric equation for the helix is 
\begin{equation} 
{\bf s}(r) =\pmatrix{\cos{i} + \tan{\psi}\cos{\phi(r)}\sin{i}  \cr \sin{\phi(r)}\tan{\psi} \cr \sin{i} - \tan{\psi}\cos{\phi(r)}\cos{i} }.
\end{equation}

where $r$ represents the distance along $\bf z'$.  The
phase angle of the helix varies according to $\phi(r) = \phi_o -
 h k r$, where $\phi_o = \phi(r = 0)$, $k= 2\pi /\lambda$ is the wave
number, and $h = -1$ for a right-handed helix. The viewing angle to
the local velocity vector along the jet is given by 
\begin{equation}
\cos{\omega(r)} = {{\bf s'}(r) \cdot {\bf x} \over |{\bf s}'(r)|}. 
\end{equation}

Substituting equation (1) into equation (2) gives 
\begin{equation}
 \cos{\omega(r)} = {\cos{i} +
\tan{\psi}\sin{i}\left[\cos{\phi(r)} + h k t \sin{\phi(r)}\right]
\over \left[\sec^2{\psi} + (h k t\tan{\psi})^2\right]^{1/2}}.
\end{equation}


For simplicity we adopt a speed of $\beta = 0.84$ along
the entire jet. The observational constraints on the fitted helix are the position of
the ridgeline on the sky, and the speed and viewing angle of the jet
at {\bf A3}.  The large jet/counter-jet flux ratio at {\bf A3} implies
that the southern jet at this location makes a smaller angle to line
of sight than the equivalent segment on the counter-jet side. The
position angle of the ridgeline at {\bf A3} then implies a right-handed helix
(i.e., the southern jet would trace a counterclockwise spiral on the
sky if it were viewed end-on). Using the same arguments, the initial phase angle
is constrained to lie in the range $-25\arcdeg < \phi_o < 25\arcdeg$.

In order to find the best combination of parameters that fit these
constraints, we calculated the predicted ridgelines of a set of models
with $20\arcdeg < i < 90 \arcdeg$, $20 < \lambda < 500$ pc, $170 <
\chi < 180$ and $2
\arcdeg < \psi < 40 \arcdeg$, and the above range of $\phi_o$. Within
this set of models, we considered only those combinations of
parameters that gave a viewing angle of $64 \pm 6 \arcdeg$ for the
inner jet. The combination that gave the best fit to the observed
ridgeline of 4C~+12.50 is given in Table~\ref{helixparms}. Although
the viewing angle of the inner jet provides a powerful constraint on
the uniqueness of this fit, the minimum in parameter space is somewhat
shallow due to the fact that the southern jet undergoes barely a
complete helical turn. Reasonably good fits to the ridgeline can be
obtained with the parameter ranges given in column 4 of
Table~\ref{helixparms}.

\begin{figure*}
\epsscale{0.9}
\plotone{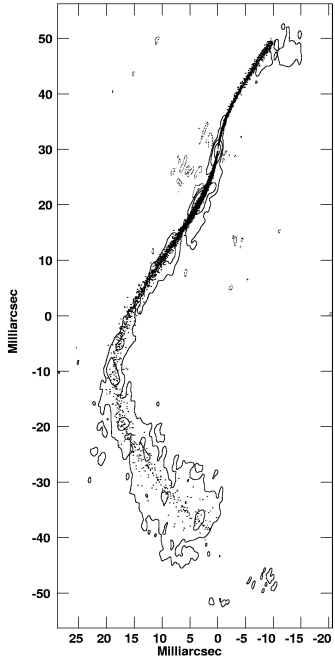}
\caption{\label{dotmodel} Sky representation of best-fit streaming jet
model, superimposed on total intensity contours for the data at $-0.4$,
0.4, 4, and 40 $\rm mJy\; beam^{-1}$. The dot density represents the
total intensity, where we have assumed a constant jet opening angle of
$1.1 \arcdeg$ and a power law emissivity decay with index of $-1.3$
along the jet.  The model for the northern jet has been truncated at
20 mas from the core. }
\end{figure*}

We show a dot-representation of our best fit model in
Figure~\ref{dotmodel}. This representation assumes a constant jet
opening angle of $1.1\arcdeg$, and a power law emissivity decay with
index of $-1.3$ along the jet. The latter value is typically observed
in quasar jets (e.g., \citealt{HOW01}). The density of the dots is
proportional to the beamed total intensity. We have truncated the
northern jet at a distance of 20 mas to match the data. Time-delays
may cause a significant difference in the apparent length of the jet
and counter-lobe, provided that $\beta\cos{i} \gtrsim 0.1$ at the end
of the jet. The predicted viewing angle at {\bf D3} in our model,
however, is $i = 88 \arcdeg$, which suggests that time delays are
unimportant, irrespective of the lobe advance speed. Although the
positions of individual features in the remainder of the jets can be
affected by time delays, the overall ridgeline will be unaffected due
to the fact that material is continually streaming along it. For these
reasons, we did not incorporate any time delay corrections into our
helical streaming model.

The model does a fairly good job at reproducing the jet ridgeline,
with the exception of the region south of {\bf A3}, where the jet
appears to curve too far to the west. There also appears to be
emission to the east of the fitted ridgeline at this location. The
other major discrepancy with the model occurs at the end of the
northern jet, where the emission appears to extend too far in a
westerly direction.  The overall fit to the total intensity is
somewhat poorer than that of the ridgeline, as the model predicts that
the counter-jet should actually increase in flux with distance from
the core as the viewing angle twists toward the line of
sight. Instead, there appears to be a gap in the counter-jet in this
region. The fact that the jet and counter-jet cannot be fit with a
symmetric model at distances further than $\sim 20$ pc from the core
suggests that the local properties of the external medium at these
radii differ considerably from one side of this source to the
other. Asymmetric environments have been invoked to explain Faraday
asymmetries in other GPS and CSS sources such as 3C 147
\citep{JSS99} and 3C 216 \citep{TGO95}.

\section{Is 4C~+12.50 a young radio source?\label{young}}

Although the extended emission associated with 4C~12.50 suggests that
it may have been active in the past, the compactness of the
parsec-scale emission suggests that the jets have only recently
restarted after a long period of inactivity. All of the age
determinations carried out for other CSOs have been based on the
relative separation speeds of components located at the extreme end of
the jet and counter-jet.  These measurements were simplified somewhat
by the fact that the terminal ``hotspots'' in these sources were quite
bright and compact, which enabled their positions to be determined
with high accuracy. In the case of 4C~+12.50, the emission at the ends
of the jet and counter-jet is too diffuse to obtain a precise lobe
advance speed. Our poor positional accuracy on component {\bf D3}
provides only an upper limit of 2 c on the apparent motion at the end
of the jet, whereas typical measured advance speeds in CSOs are on the
order of 0.3 c \citep{OCP99}. The latter speed would imply an age of
only $\sim 1700$ y for the parsec-scale structure in 4C~+12.50. A slow
lobe advance speed would also imply a significant deceleration of the
jet from its initial value of $\beta = 0.84$ in the inner jet. This
deceleration could conceivably occur either at the strong bend
(component {\bf D2}), or at component {\bf D3}. \cite{SNB98} have
incorporated such jet deceleration in a model of GPS sources that can
explain their symmetric morphologies and flux outburst behavior.

The overall host galaxy properties of IRAS 13451+1232 are generally
consistent with the notion that the parsec-scale outflow is fairly
recent. ULIGs are thought to represent an early stage of
quasar evolution \citep{SM96}, and those with relatively ``warm''
infrared colors such as IRAS 13451+1232 may be in a transition from a
cold ULIG phase to an optical quasar phase
\citep{SSE88}. \cite{M89} proposes that IRAS 13451+1232
 is a young progenitor of a bright radio IR spiral such as Arp~220,
 and is still confined by large densities of gas and dust. It was also
 the most morphologically disturbed object in a sample of nearby
 luminous radio galaxies studied by \cite{SH89}. According to
\cite{GS86}, the disturbed outer isophotes imply that the merger
was both violent and recent, since phase mixing should have restored
symmetry in about $10^{8-9}$ y. 

Although the merger timescales for massive galaxies (on the order of
$10^8$ y) are much longer than the observed ages of CSOs, the
timescale for black hole mergers which could trigger the radio jets are
likely much shorter (D. Meier, private communication). After
such a merger, the spin axis of the black hole is predicted to align
with the accretion disk in a relatively short time of about $10^5$ y
\citep{NP98}, which would dampen any precession of the radio
jets. The prevalence of ``S'' shaped morphologies in CSOs (e.g.,
0108+388, 3C 186, 3C 196, 1946+708, and 2352+495) may simply reflect
the fact that their black hole spin axes are still precessing, and
have not had sufficient time to stabilize. More detailed observations
of the extended emission associated with GPS and CSO sources such as
4C~12.50 could prove useful for constraining the direction of the radio
jet axis during periods of previous activity.

\section{Summary \label{summary}}

We summarize the main results of our multi-epoch VLBA study of the
kinematics and polarization properties of the compact symmetric object
4C~+12.50 as follows:

1. 4C~+12.50 has a relatively low integrated linear polarization on
parsec-scales ($m = 0.7\%$) that is typical of other CSO and GPS
sources, and is expected given the large amount of gas present in the
host galaxy. The high-resolution polarization images, however, reveal
isolated regions in the southern jet with exceedingly high fractional
polarizations that range up to $m = 60\%$. These values are highly
unusual for a CSO, and suggest that the depolarizing medium is
somewhat clumpy. These findings illustrate the need for more
high-resolution polarization-sensitive observations of other CSO and
GPS sources in order to better understand their jet magnetic field
properties and local environments.

2. A kinematic analysis of 4C~+12.50 reveals jet component 
speeds of $1.0 \pm 0.3$ and $1.2 \pm 0.2$ c, which represent the first
positive detections of superluminal motion in a compact symmetric
object. We are able to fit both the apparent motion
and the observed ``$\varepsilon$'' shape of the source using a
conical helical jet. The helical ridgeline is likely the result of
small, periodic perturbations at the jet nozzle that are amplified by 
Kelvin-Helmholtz instabilities in the flow. These perturbations could
arise from orbital motion of the black hole and/or precession
of the jet nozzle. The best-fit model suggests that the nozzle is
precessing around a cone with half-angle $23 \arcdeg$, whose axis lies
at an angle of $82 \arcdeg$ to the line of sight. The jet has a bulk
flow speed of 0.84 c, and follows a right-handed helical path with a
wavelength of 280 pc.

3. The detection of relativistic speeds in the jet of 4C~+12.50
implies the existence of a beamed counterpart population to the CSOs
that has yet to be identified. Although the properties of these
objects are not expected to be as extreme as blazars, they should have
have high apparent flux densities, and are likely already present
in current radio AGN catalogs.

\acknowledgments
 This research  has made use of data from the following sources:

 The University of Michigan Radio Astronomy Observatory, which
 is supported by the National Science Foundation and by funds from the
 University of Michigan.

The NASA/IPAC Extragalactic Database (NED) which is operated by the
Jet Propulsion Laboratory, California Institute of Technology, under
contract with the National Aeronautics and Space Administration.

\begin{deluxetable}{lllccl} 
\tablecolumns{6} 
\tablewidth{0pt}  
\tablecaption{\label{journal}Journal of 15 GHz Observations}
\tablehead{\colhead{Observing} &  &
\colhead{Polarization} & \colhead{Bandwidth\tablenotemark{a}} &
\colhead{Integration} & \\
\colhead{Date}&\colhead{Array} &   \colhead{Mode} &\colhead{[MHz]}
&\colhead{Time [min]} &\colhead{Reference} }
\startdata 	 
1996 Apr 22 & VLBA+Y1 & Dual Circular & 16 & 110 & \cite{SDO01} \\
1997 Aug 28 & VLBA  & LL only & 64 &40 &\cite{ZRK02} \\
1998 Nov 1  & VLBA & LL only & 64 &40& \cite{ZRK02}  \\
1999 Nov 6  & VLBA & LL only & 64 &200& \cite{ZRK02} \\
2001 Jan 4  & VLBA+Y1 & Dual Circular & 32 & 540& This work \\
\enddata 
\tablenotetext{a}{Observing bandwidth per hand of circular polarization} 
\end{deluxetable} 

\begin{deluxetable}{lrrr} 
\tabletypesize{\small}
\tablecolumns{4} 
\tablewidth{0pt}  
\tablecaption{\label{cptpos} Model Component Fits}  
\tablehead{\colhead{Cpt.} & \colhead {$r$} &   \colhead{$\theta$} & \colhead{S}  \\  
\colhead{Name} & \colhead{[mas]} & \colhead{[deg.]} & \colhead{[mJy]} \\  
\colhead{(1)} & \colhead{(2)} & \colhead{(3)} & \colhead{(4)}   
 } 
\startdata 
\sidehead{1996 Apr 22} 
A1 & \n &   \n  &  203  \\* 
A2 &  0.9 $\pm 0.1$ &$   169 $ &   59  \\* 
A3 &  1.7 $\pm 0.1$&$   160 $ &   30  \\* 
B1 &  7.6 $\pm 0.3$&$   169 $ &   14  \\* 
B2 &  9.7 $\pm 0.2$&$   162 $ &  224  \\* 
C & 20.9 $\pm 0.2$&$   157 $ &   87  \\* 
D1 & 44.5 $\pm 0.1$&$   155 $ &  161  \\* 
D2 & 52.2 $\pm 0.2$&$   161 $ &   92  \\* 
D3 & 65.9 $\pm 0.6$&$   177 $ &  139  \\* 
\sidehead{1997 Aug 28} 
A1 & \n    &     \n  &  170  \\* 
A2 &  0.9 $\pm 0.1$&$   171 $ &   74  \\* 
A3 &  1.8 $\pm 0.1$&$   165 $ &   38  \\* 
B1 &  6.7 $\pm 0.3$&$   171 $ &    6  \\* 
B2 &  9.8 $\pm 0.2$&$   162 $ &  225  \\* 
C & 19.1 $\pm 0.2$&$   155 $ &   91  \\* 
D1 & 44.5 $\pm 0.1$&$   155 $ &  146  \\* 
D2 & 52.6 $\pm 0.2$&$   162 $ &   83  \\* 
D3 & 66.4 $\pm 0.6$&$   177 $ &   89  \\* 
\sidehead{1998 Nov 1} 
A1 & \n &     \n  &  159  \\* 
A2 &  1.0 $\pm 0.1$&$   170 $ &   64  \\* 
A3 &  2.0 $\pm 0.1$&$   166 $ &   24  \\* 
B1 &  7.1 $\pm 0.3$&$   172 $ &    7  \\* 
B2 &  9.9 $\pm 0.2$&$   162 $ &  196  \\* 
C & 20.9 $\pm 0.2$&$   156 $ &   80  \\* 
D1 & 44.5 $\pm 0.1$&$   154 $ &  143  \\* 
D2 & 52.3 $\pm 0.2$&$   161 $ &   74  \\* 
D3 & 65.9 $\pm 0.6$&$   177 $ &   85  \\* 
\sidehead{1999 Nov 6} 
A1 &  \n &     \n  &  169  \\* 
A2 &  1.2 $\pm 0.1$&$   168 $ &   61  \\* 
A3 &  2.3 $\pm 0.1$&$   164 $ &   20  \\* 
B1 &  6.9 $\pm 0.3$&$   172 $ &    8  \\* 
B2 &  9.8 $\pm 0.2$&$   162 $ &  194  \\* 
C & 20.9 $\pm 0.2$&$   156 $ &   91  \\* 
D1 & 44.5 $\pm 0.1$&$   155 $ &  155  \\* 
D2 & 52.4 $\pm 0.2$&$   161 $ &   85  \\* 
D3 & 66.0 $\pm 0.6$&$   178 $ &   97  \\* 
\sidehead{2001 Jan 4} 
A1 &  \n &     \n  &  183  \\* 
A2 &  1.3 $\pm 0.1$&$   165 $ &   41  \\* 
A3 &  2.3 $\pm 0.1$&$   165 $ &   11  \\* 
B1 &  7.1 $\pm 0.3$&$   173 $ &   13  \\* 
B2 &  9.9 $\pm 0.2$&$   161 $ &  155  \\* 
C & 21.0 $\pm 0.2$&$   156 $ &   80  \\* 
D1 & 44.5 $\pm 0.1$&$   155 $ &  151  \\* 
D2 & 52.4 $\pm 0.2$&$   161 $ &   79  \\* 
D3 & 66.2 $\pm 0.6$&$   177 $ &  124  \\* 
\enddata 
\tablecomments{Columns are as follows: (1) Component name; (2) Distance 
from core, in milliarcseconds; (3) Position angle with respect to 
core; (4) Flux density in mJy.} 
\end{deluxetable} 

\begin{deluxetable}{lrrrll} 
\tablecolumns{6}
\tablewidth{0pt}  
\tablecaption{\label{cptprops}Jet Component Properties}  
\tablehead{\colhead{Cpt.}  & \colhead{Maj.} &  	\colhead{Axial} &
\colhead{PA} &  \colhead{$\mu$} & \\  
\colhead{Name}  & \colhead{Axis} & \colhead{Ratio} & \colhead{[deg.]}
	&\colhead{(mas/y)} & \colhead{$\beta_{app} $} \\  
\colhead{(1)} & \colhead{(2)} & \colhead{(3)} & \colhead{(4)} &  
\colhead{(5)} & \colhead{(6)} } 
\startdata 
A1  & 0.47 & 0.67 & $   -31 $ &\n &\n  \\* 
A2 &  0.52 & 0.13 & $   -60 $& $0.13 \pm 0.04$\tablenotemark{a}  & $1.0
\pm 0.3$\\* 
A3  &  0.79 & 0.39 & $    43 $ & $0.15 \pm 0.03$ & $1.2 \pm 0.2$ \\* 
B1  & 0.00 & \n &     \n  & $0.09 \pm 0.1$\tablenotemark{a} & $0.7
\pm 0.8$  \\* 
B2  & 2.32 & 0.34 & $   -33 $ & $0.03 \pm 0.05$ & $0.2 \pm 0.4$ \\* 
C  &5.86 & 0.22 & $   -16 $ & $<0.1$ & $<0.8$ \\* 
D1  & 6.42 & 0.39 & $     8 $& $<0.03$ & $<0.2$  \\* 
D2  & 4.73 & 0.61 & $    24 $& $<0.12$ & $<1.0$ \\* 
D3  & 7.41 & 0.47 & $     6 $ & $<0.3$ & $<2$  \\* 
\enddata 
\tablenotetext{a}{Calculated excluding the 1996 Apr 22 epoch}
\tablecomments{Columns are as follows: (1) Component name;  (2) Major axis of fitted component, in 
milliarcseconds; (3) Axial ratio of fitted component; (4) Position 
angle of component's major axis; (5) Component motion in mas/yr; (6)
Component speed in units of c.} 
\end{deluxetable}

\begin{deluxetable}{lcccc} 
\tablecolumns{5} 
\tablewidth{0pt}  
\tablecaption{\label{helixparms}  Helical Streaming Model Parameters}  
\tablehead{\colhead{Parameter} & \colhead {Symbol} &   \colhead{Best
Fit Value}
 &\colhead{Range} & \colhead{Units}   } 
\startdata 

 Half-opening angle of streaming cone & $\psi$  & \phn  23& $\pm 2$ & deg.\\* 
 Sky position angle of cone axis & $\chi$ & 175 & $\pm 2$ & deg.\\*
 Viewing angle to cone axis & $i$ & \phn 82 & $80 - 87$ & deg. \\*
 Initial phase angle &  $\phi_o$ &\phn\phn   5 &$\pm 6$ & deg.\\*
 Helix wavelength  & $\lambda$ & 280 &  $270 - 300$  & pc \\*
 Helix handedness &$h$ &  $-1$ (right) & \n & \n  \\*
\enddata
\end{deluxetable}

\end{document}